\documentclass[10pt,conference]{IEEEtran}

\IEEEoverridecommandlockouts
\usepackage{cite}
\usepackage{amsmath,amssymb,amsfonts}
\usepackage{algorithmic}
\usepackage{graphicx}
\usepackage[ruled,vlined,linesnumbered]{algorithm2e}
\usepackage{amsthm}
\usepackage{subfigure}

\usepackage{textcomp}
\usepackage{xcolor}
\usepackage{xspace}
\def\BibTeX{{\rm B\kern-.05em{\sc i\kern-.025em b}\kern-.08em
    T\kern-.1667em\lower.7ex\hbox{E}\kern-.125emX}}

\newcommand{\tabincell}[2]{\begin{tabular}{@{}#1@{}}#2\end{tabular}}

\usepackage[framemethod=tikz]{mdframed}
\newcounter{finding}
\newcommand{\finding}[1]{\refstepcounter{finding}
  \vspace{2.3mm}
 \begin{mdframed}[linecolor=gray,roundcorner=12pt,backgroundcolor=gray!15,linewidth=3pt,innerleftmargin=2pt, leftmargin=0cm,rightmargin=0cm,topline=false,bottomline=false,rightline = false]
  \textbf{Ans. to RQ\arabic{finding}:} #1
 \end{mdframed}
 \vspace{2.3mm}
}
\usepackage{url}
\usepackage{multirow}

\usepackage{tikz}
\usepackage{colortbl}
\usepackage{array}

\begin{document}

% \title{Conference Paper Title*\\

% \usepackage{xspace}
\newcommand{\toolName}{\textit{SateLight}\xspace}

\newcommand{\compareOne}{\textit{PT4Cloud}\xspace}
\newcommand{\compareTwo}{\textit{Metior}\xspace}
\newcommand{\wjf}[1]{{\color{blue}{#1}}}

% \title{\toolName: Performance Testing for Serverless Computing\vspace{-2em}}

\title{\toolName: A Satellite Application Update Framework for Satellite Computing}

\makeatletter
\newcommand{\linebreakand}{%
  \end{@IEEEauthorhalign}
  \hfill\mbox{}\par
  \mbox{}\hfill\begin{@IEEEauthorhalign}
}
\makeatother

\author{
  \IEEEauthorblockN{Jinfeng Wen}%
\IEEEauthorblockA{\textit{Beijing University of }\\ \textit{Posts and Telecommunications} \\
Beijing, China\\
jinfeng.wen@bupt.edu.cn}%
  \and
  \IEEEauthorblockN{Jianshu Zhao}%
\IEEEauthorblockA{\textit{Beijing University of }\\ \textit{Posts and Telecommunications} \\
Beijing, China\\
jianshuzhao@bupt.edu.cn}%
  \and
 \IEEEauthorblockN{Zixi Zhu}%
\IEEEauthorblockA{\textit{Beijing University of }\\ \textit{Posts and Telecommunications} \\
Beijing, China\\
zhuzixi.zzx@gmail.com}%
  \linebreakand % <------------- \and with a line-break
  \IEEEauthorblockN{Xiaomin Zhang}%
\IEEEauthorblockA{\textit{Beijing University of }\\ \textit{Posts and Telecommunications} \\
Beijing, China\\
zxm987626@bupt.edu.cn}%
  \and
  \IEEEauthorblockN{Qi Liang}%
\IEEEauthorblockA{\textit{Beijing University of }\\ \textit{Posts and Telecommunications} \\
Beijing, China\\
liangqi@bupt.edu.cn}%
      \linebreakand % <------------- \and with a line-break
  \IEEEauthorblockN{Ao Zhou{*}\thanks{{*}Corresponding author.}}%
\IEEEauthorblockA{\textit{Beijing University of }\\ \textit{Posts and Telecommunications} \\
Beijing, China\\
aozhou@bupt.edu.cn}%
  \and
  \IEEEauthorblockN{Shangguang Wang}%
\IEEEauthorblockA{\textit{Beijing University of }\\ \textit{Posts and Telecommunications} \\
Beijing, China\\
sgwang@bupt.edu.cn}
}

\maketitle

% \vspace{-10mm}

\begin{abstract}

Satellite computing is an emerging paradigm that empowers satellites to perform onboard processing tasks (i.e., \textit{satellite applications}), thereby reducing reliance on ground-based systems and improving responsiveness. However, enabling application software updates in this context remains a fundamental challenge due to application heterogeneity, limited ground-to-satellite bandwidth, and harsh space conditions. Existing software update approaches, designed primarily for terrestrial systems, fail to address these constraints, as they assume abundant computational capacity and stable connectivity.

To address this gap, we propose \toolName, a practical and effective satellite application update framework tailored for satellite computing. \toolName leverages containerization to encapsulate heterogeneous applications, enabling efficient deployment and maintenance. \toolName further integrates three capabilities: (1) a content-aware differential strategy that minimizes communication data volume, (2) a fine-grained onboard update design that reconstructs target applications, and (3) a layer-based fault-tolerant recovery mechanism to ensure reliability under failure-prone space conditions. Experimental results on a satellite simulation environment with 10 representative satellite applications demonstrate that \toolName reduces transmission latency by up to 91.18\% (average 56.54\%) compared to the best currently available baseline. It also consistently ensures 100\% update correctness across all evaluated applications. Furthermore, a case study on a real-world in-orbit satellite demonstrates the practicality of our approach.

% Satellite computing is an emerging paradigm that enables satellites to perform onboard processing tasks, reducing reliance on ground-based systems. However, realizing software updates in satellite environments remains a major challenge due to application heterogeneity, limited onboard resources, bandwidth-constrained communication, and the harsh conditions of space. Existing software update approaches are insufficient to address these constraints, as they assume abundant computational resources and stable connectivity.

% To bridge this gap, we propose \toolName, a practical and efficient software update framework tailored for satellite computing. \toolName adopts containers to encapsulate heterogeneous satellite applications, enabling lightweight deployment, maintenance, and isolation. \toolName further integrates the three key capabilities: a content-aware differential transmission strategy to minimize uplink data volume, a fine-grained onboard update mechanism with minimal runtime overhead, and a fault-tolerant recovery to ensure update reliability under failure-prone conditions. We evaluate \toolName on a satellite simulation platform across a diverse set of satellite applications implemented in different programming languages. Experimental results demonstrate that \toolName significantly reduces application transmission latency by up to XX\%, compared to baseline methods. Moreover, we validate the practicality of \toolName through a case study on a real-world on-orbit satellite.

\end{abstract}


\section{Introduction}\label{sec:introduction}

% Satellite computing is an emerging paradigm, which empowers satellites with computing resources to support on-orbit task processing. Advanced computing platforms onboard can be transformed as sophisticated data processing infrastructure and enable tasks in space as they are processed today on the ground. Such tasks can be regarded as satellite application software, which can capture and process the raw data, identify the features of interest, and return the interesting results.

Satellite computing~\cite{wang2023satellite, zhang2024resource} is an emerging paradigm that endows Low Earth Orbit (LEO) satellites with onboard computing capabilities, enabling them to process mission tasks directly in orbit. LEO satellites~\cite{ren2024sateriot, ouyang2023joint}, exemplified by leading constellations such as Telesat, OneWeb, and SpaceX, are undergoing rapid development~\cite{del2019technical}. Operating at orbital altitudes between 500 km and 1,000 km, these satellites have traditionally been limited to data collection and transmission.
% LEO satellites, as demonstrated by prominent constellations Telesat, OneWeb, and SpaceX, are rapidly evolving~\cite{del2019technical}. They operate at orbital altitudes of 500 km to 1,000 km and are traditionally limited to data collection and transmission. 
However, growing demands for real-time responsiveness and massive space-borne data have driven a shift towards autonomous, intelligent satellites with integrated processing capabilities. Modern satellites are now equipped with processors, memory, and accelerators, forming onboard infrastructure capable of executing complex tasks, i.e., \textit{satellite applications}.

% and have recently demonstrated the feasibility of providing global access to ground-based satellites via space links. Traditionally, LEO satellites primarily acted as data collectors, transmitting raw data back to the ground for analysis. However, with the growing demand for real-time responsiveness and the increasing volume of spaceborne data, the role of satellites has shifted towards becoming autonomous data processing platforms. These satellites are equipped with powerful processors, memory, and AI accelerators, forming sophisticated onboard computing infrastructure.

Satellite applications refer to the programs deployed on satellites to perform mission-specific computations, which can capture raw sensor data, analyze features of interest, and make decisions or transmit results back to the ground stations. Representative satellite applications~\cite{xing2024deciphering, xing2023earth, pfandzelter2024komet, zhai2024seco, liu2024orbit} include image encoding, object detection, feature tracking, etc. However, once the satellite is launched, the functionalities of satellite applications are generally fixed, rendering it inflexible to accommodate evolving mission objectives or unforeseen requirements. This lack of adaptability poses a significant limitation in the dynamic and long-duration context of modern space missions. Thus, the capability to remotely update satellite applications from ground stations to satellites, whether for new features or bug fixes, has emerged as a critical and urgent requirement in satellite computing.

% and cannot be flexibly adapted to evolving mission requirements. In the dynamic context of space missions, this significantly constrains the adaptability and long-term utility of satellites. \wjf{Adding new features or fixing known detects are important maintenance tasks. Therefore, the ability to update satellite application software after launch has become an urgent and critical need.}

% Existing software updates-related research has been explored in ground-based systems, dividing into pre-update, during the update, and post-update processes. For example, XX, XX, XXX~\cite{}. The most-related studies pre-update stage and during the update.  However, pre-update approaches have primarily focused on recommending what to update, without addressing how to execute the actual process from the old version to the new one. During the update execution phase, dynamic software techniques often rely on language-specific static analysis, runtime monitoring, and large test cases to determine sate update points. .....

Despite its critical importance, effective application updating in satellite computing remains an unsolved problem. Existing software update approaches in terrestrial environments are generally divided into three categories: pre-update, in-update, and post-update. Pre-update approaches have focused on recommending \textit{what} to update~\cite{di2016would, zhou2020user, hassan2018rudsea, liu2025automatically}, assuming updates can be directly applied through full replacements. In-update studies have involved dynamic software updates, which enable runtime changes without service interruption~\cite{zhao2014automated, cazzola2016dodging, giuffrida2016automating}. However, these methods have depended on heavyweight static analysis, language-specific instrumentation, and extensive test suites, which could make them unsuitable for resource-constrained satellite systems. Post-update research has explored user feedback and update behavior analysis~\cite{mirhosseini2017can, saidani2022tracking, di2022software, kotzias2019mind}, but focused on software evolution rather than specific update mechanisms.

Updating satellite applications introduces several unique and under-explored challenges. (1) \textit{Heterogeneous applications and constrained resources.} Satellite systems host diverse software stacks, making language-specific update approaches infeasible. Limited onboard resources (e.g., CPU, memory, and energy) further restrict complex update logic. (2) \textit{Low-bandwidth, intermittent communication.} Communication uplink capacity from the ground to the satellite is severely limited (tens to hundreds of kbps), with short, infrequent contact windows, leading to a high risk of incomplete or delayed application update delivery. (3) \textit{Harsh and unreliable space environments.} Radiation and thermal shifts in space increase the likelihood of in-orbit faults, turning failed updates into potentially mission-critical events. To the best of our knowledge, limited research has addressed these challenges. A critical gap remains in designing satellite application update approaches that can effectively handle application heterogeneity, communication constraints, and reliability requirements.

To address these challenges, we present \toolName, a practical and effective satellite application update framework for satellite computing. At its core, \toolName adopts a container-based design to encapsulate heterogeneous satellite applications. This abstraction leverages the well-established benefits of containers (e.g., flexibility and isolation), thereby simplifying application deployment and maintenance in resource-constrained satellite systems. Containerization also mitigates the complexity of language-specific software update logic, enabling uniform update management based on containers.

Built on this foundation, \toolName introduces three core capabilities: (1) a communication-efficient upload strategy that minimizes uplink bandwidth usage through content-aware differential identification; (2) a fine-grained onboard update strategy that enables deterministic reconstruction of target application versions; and (3) a fault-tolerant recovery mechanism that ensures update reliability by leveraging the layered nature of containers for rapid recovery upon failure. Specifically, \toolName employs a content-aware differential algorithm to extract semantic deltas between the existing and updated application versions. These deltas are encoded and transmitted to the satellite, reducing the data transmission overhead. Upon reception, the onboard satellite system performs an application reconstruction by applying fine-grained updates at line or chunk granularity to the original containerized application. To guarantee update reliability, \toolName integrates a layer-based rollback mechanism: upon detecting update failures, it instantly reverts to the last stable container layer, thereby eliminating the need for a full re-upload.

We implement \toolName and evaluate its effectiveness using 10 representative satellite applications written in diverse programming languages, within a satellite simulation environment. Results show that \toolName reduces uplink transmission latency by an average of 99.99\%, 73.06\%, and 56.54\% compared to three baselines. Despite adopting a fine-grained onboard application reconstruction strategy, \toolName incurs only a negligible overhead, increasing execution time by around 2 seconds on average. Moreover, it consistently achieves 100\% update correctness across all evaluated applications.
% Building on this transmission advantage, \toolName further improves end-to-end update performance by an average of \wjf{XX\%, XX\%, and XX\%} over the same baselines.
% Results show that \toolName reduces the uplink transmission latency by an average of \wjf{XX\%, XX\%, and XX\%} compared to three baseline approaches. Based on transmission benefit, \toolName improvement end-to-end update performance by an average of \wjf{XX\%, XX\%, and XX\%}, compared to three baseline approaches.
We validate the practicality of \toolName through a case study on BUPT-2, a real-world in-orbit satellite equipped with onboard cloud-native computing. We release a public repository~\cite{ourdata} containing data and code used in this study to support reproducibility and future research.

\section{Background}\label{sec:background}

% This section provides an overview of the satellite-ground system architecture and outlines key challenges associated with performing satellite application updates.

\subsection{Satellite-Ground System Architecture}

% Due to orbital constraints, opportunities for direct communication between the LEO satellite and ground systems are highly time-constrained and intermittent. Generally, the satellite establishes 4 to 6 contact windows per data, each lasting approximately 10 minutes~\cite{tao2023transmitting}. Moreover, the connection between the satellite and the ground can be affected by many factors, such as weather, satellite orientation, and relative antenna position between the radios and the satellites. This short-duration, periodic, and vulnerable connectively imposes stringent requirements. Data transmission and command uplink/downlink operations must be completed with high efficiency within the limited communication windows.

% Due to orbital constraints, direct communication between LEO satellites and ground stations is highly intermittent and time-limited. Generally, a satellite establishes 4 to 6 contact windows per day, each lasting approximately 10 minutes~\cite{tao2023transmitting}. Moreover, the availability of these links sometimes is susceptible to various factors, including weather conditions, satellite orientation, and the relative alignment of ground and onboard antennas. This short-duration, periodic, and fragile connectivity imposes stringent requirements, necessitating efficient data transmission and uplink/downlink operations within narrow communication windows. In our scenario, the revised applications are uploaded via the uplink operation to the satellite to execute the corresponding functionalities.

\begin{figure}[t]
	\centering
    \includegraphics[width=0.4\textwidth]{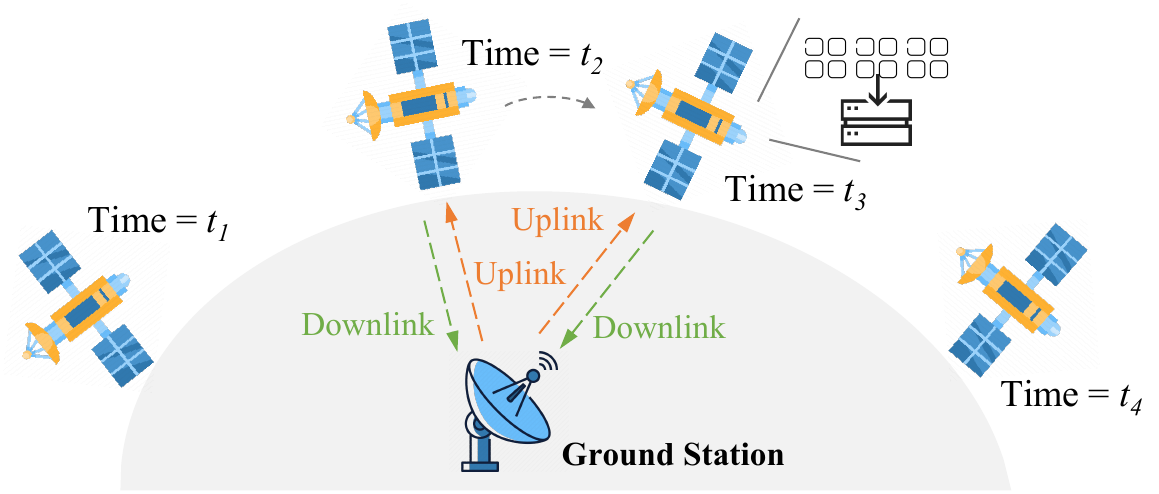}
    % \vspace{-4mm}
    \caption{Satellite-ground communication architecture.}
    \label{fig:satellite-ground}
    \vspace{-3mm}
\end{figure}

As illustrated in Fig.~\ref{fig:satellite-ground}, the interaction between LEO satellites and ground stations follows a highly time-sensitive and intermittent communication model, governed by orbital mechanics and visibility constraints. For instance, at $t_1$, the satellite is approaching the visibility zone but remains outside the effective communication range. At $t_2$, it enters the communication window, enabling both uplink (e.g., command transmissions, application updates) and downlink (e.g., telemetry, sensor data) operations. At $t_3$, the satellite exits the communication range and autonomously executes onboard tasks. At $t_4$, it passes beyond the ground station's reach and awaits the next contact opportunity.

% As shown in Fig.~\ref{fig:satellite-ground}, the satellite-ground interaction proceeds through the following stages. For example, at Time = $t_1$, the satellite is approaching the ground station’s visibility zone but remains outside the communication range. At Time = $t_2$, the satellite enters the communication window, enabling both uplink (e.g., application updates, commands) and downlink (e.g., telemetry, collected data) operations. At Time = $t_3$, after exiting the visibility zone, the satellite continues its orbit and autonomously executes the received instructions. At Time = $t_4$, the satellite has moved further beyond the ground station’s coverage and is again outside the communication range, awaiting the next contact opportunity.

Communication with satellites is intermittent, with typical contact windows occurring 4–6 times per day, each lasting approximately 10 minutes~\cite{tao2023transmitting}. Moreover, uplink bandwidth generally ranges from tens to a few hundred kbps~\cite{tao2023transmitting, devaraj2019planet, vasisht2021l2d2, wang2023satellite}. These impose strict requirements on data transmission to ensure reliable operations within each limited contact window.

% Due to orbital constraints and satellite mobility, direct communication between LEO satellites and ground stations on Earth is intermittent and time-constrained, generally limited to 4–6 contact windows per day, each lasting approximately 10 minutes~\cite{tao2023transmitting}. The availability of these links is further affected by atmospheric conditions, satellite orientation, and antenna alignment, resulting in periodic and fragile connectivity. Furthermore, the communication bandwidth is constrained by the physical limitations of long-range space-to-ground links. This connectivity pattern imposes stringent requirements on data prioritization and transmission efficiency to ensure reliable mission execution within each limited contact period.

\subsection{Key Challenges in Satellite Application Updates}

% Realizing efficient application updates in satellite systems involves overcoming several challenges arising from the unique constraints of the space environment. These challenges can be categorized into the following three aspects.

Ensuring effective and reliable satellite application updates is non-trivial due to the following challenges:

$\bullet$ \textit{Challenge 1: Application Heterogeneity and Onboard Resource Constraints.} Satellites can host diverse applications built using different programming languages, e.g., C, C++, and Python. This application heterogeneity complicates the design of a unified software update mechanism. Moreover, satellite systems operate under stringent resource constraints, including limited CPU cycles, memory, and energy. These limitations preclude the use of heavyweight software update methods~\cite{zhao2014automated, cazzola2016dodging, giuffrida2016automating} (e.g., complex program analysis or large test cases). The limited storage capacity of satellites further constrains the retention of multiple versions of applications.

% Storage limitations of satellites also restrict the ability to retain multiple historical application versions.

% Satellites host diverse applications developed in various programming languages and platforms, creating a fragmented software ecosystem. Supporting customized application update strategies for each is impractical. Meanwhile, the constrained onboard resources, including CPU, memory, storage, and energy, pose further challenges. On one hand, running heavyweight update logic is an inability solution, such as complex program analysis or large test cases. On the other hand, the satellite has limited storage, which restricts the retention of multiple application versions or recovery states. 

$\bullet$ \textit{Challenge 2: Limited Communication Bandwidth.} Uplink channels in LEO satellites are fundamentally constrained by low bandwidth and short transmission windows with ground stations. Large application updates can easily exceed the available bandwidth within a single pass, leading to partial transmissions and incomplete deployments. These constraints place greater demands on maximizing communication efficiency to ensure the delivery of critical missions.
% This vulnerability is exacerbated by the lack of persistent connectivity, necessitating mechanisms that can support incremental or resumable delivery.

% Satellite communication links suffer from inherently low uplink bandwidth, generally limited to tens or hundreds of kbps. This constraint leads to high transmission latency, delaying the delivery of new application versions, especially for large updates, and vulnerability to interruption, as communication windows are short and transmission sessions can be easily disrupted.

$\bullet$ \textit{Challenge 3: Space Environment Risk in Orbit.} The space environment is subject to radiation, extreme temperature fluctuations, and mechanical stress, which could pose threats to application updates. Any failure in update execution or application logic can compromise mission-critical functions, and in extreme cases, lead to permanent satellite malfunction. Consequently, satellite systems impose stringent reliability requirements on any update mechanism.

\section{\toolName: Application Update Framework}

\begin{figure*}[t]
	\centering
    \includegraphics[width=0.97\textwidth]{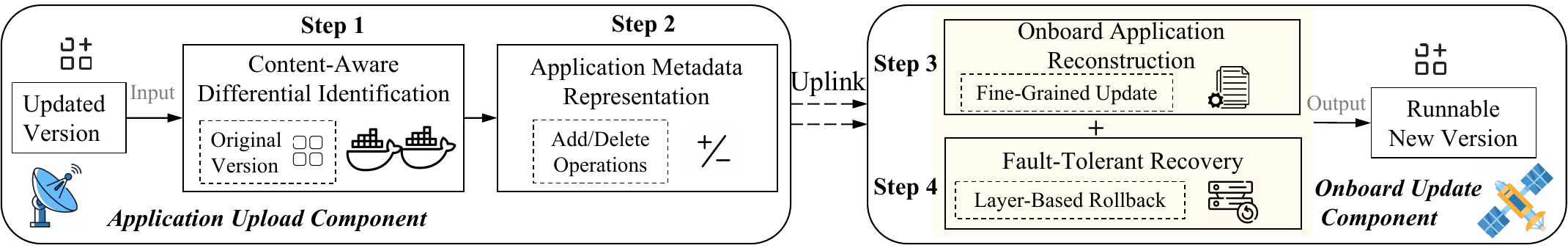}
    % \vspace{-3mm}
    \caption{The satellite application update workflow of \toolName.}
    \label{fig:overview}
    \vspace{-3mm}
\end{figure*}

To systematically address the challenges of application updates in satellite-ground systems, we present \toolName, a unified framework that enables a lightweight, modular, and reliable software update tailored for satellite computing. The design of \toolName is guided by the following three principles.
% : abstraction for heterogeneity, minimization for transmission, and reliability guarantee for uncertain conditions.

% we propose \toolName, a unified framework that combines container-based application encapsulation with content-aware intelligent updates. 
 
% These principles enable a lightweight, modular, and reliable software update tailored for satellite computing.

% To address the key challenges of software maintenance and evolution in satellite-ground systems, we propose the design of our software update framework named \toolName. This framework is both container-based and context-aware, aiming to enable efficient, lightweight, and reliable application updates under the stringent conditions of satellite systems. 

\textbf{\textit{Design Principle for Challenge 1:}} \textit{Abstraction to Unify Heterogeneity.} Satellite applications span multiple languages and different configurations, complicating update mechanisms and maintenance logic. Rather than attempting to generalize over heterogeneous toolchains, \toolName sidesteps this complexity through application containerization as a unifying abstraction. Each application is encapsulated in a self-contained, portable unit that includes all dependencies (runtime, libraries, configurations), enabling uniform deployment and update workflows. This decouples update logic from application internals, reducing per-application engineering overhead. 
% Moreover, the containers offer built-in resource isolation and modular lifecycle management, which is critical for low-resource onboard environments.

% To tackle the challenge of software heterogeneity and onboard resource limitations, \toolName employs the containerized design to abstract application dependencies and runtime environments. This enables a unified, modular deployment model capable of supporting diverse heterogeneous satellite applications without the need for application-specific update mechanisms. Each container image~\cite{docker} encapsulates all necessary components, e.g., code, runtime, libraries, system tools, and configurations, into a lightweight, portable, and executable unit. By isolating satellite applications within containers, \toolName ensures general, consistency and minimal disruption during updates.

\textbf{\textit{Design Principle for Challenge 2:}} \textit{Differentiation to Minimize Bandwidth Load.} Recognizing the limitations of uplink communication bandwidth, \toolName adopts a content-aware differential upload strategy. Instead of transmitting the complete application, \toolName analyzes and separates fine-grained semantic code changes between application versions, not file types. \toolName identifies these changes at the level of code lines or chunks.
These differences are packaged into compact update payloads, reducing data volume and enabling efficient delivery within short communication windows.
% These fine-grained differences are then packaged into compact upload data, which reduces data volume and makes update delivery feasible within short and intermittent communication windows.

% To address the challenge of limited uplink communication bandwidth, \toolName introduces a context-aware update strategy that reduces transmission overhead. It identifies and extracts only the \textit{changed context}, i.e., the set of code differences between the new and original application versions. Then, it transmits this minimal set as a lightweight update package. This design ensures that only essential data is sent over the bandwidth-limited satellite uplink, improving update efficiency and reducing the risk of interruption during transmission.

\textbf{\textit{Design Principle for Challenge 3:}} \textit{Reliability Though Reversible Execution.} Space environments are sometimes unreliable, with radiation and power fluctuations posing risks to the update process. \toolName incorporates a fault-tolerant recovery mechanism. Instead of relying on fine-grained state reconstruction or re-downloading the entire application, \toolName leverages the inherent layered structure of containers, enabling it to rapidly track checkpoints for each application. 

Guided by the aforementioned design principles, as illustrated in Fig.~\ref{fig:overview}, \toolName implements an application update framework for satellite-ground systems through two core components. The \textit{Application Upload Component}, deployed on the ground, is responsible for performing content-aware differential analysis between the updated and original containerized application versions (Step 1). It identifies semantic changes and encodes them into an expressive metadata representation (Step 2). The \textit{Onboard Update Component}, residing in the satellite system, receives the update package and executes a fine-grained application reconstruction process to ensure update correctness (Step 3). It also integrates a fault-tolerant recovery mechanism based on container layering, enabling rapid application rollback in case of update failures (Step 4).

\subsection{Application Upload Component}

% This component takes an updated version of the containerized application as input and generates an optimized update package as output. It primarily consists of context-aware difference identification and application metadata representation.

This component operates in two main stages: Content-Aware Differential Identification and Application Metadata Representation, as depicted in Step 1 and Step 2 of Fig.~\ref{fig:overview}.

\textit{\textbf{Step 1: Content-Aware Differential Identification.}}  In this phase, \toolName conducts a hierarchical comparison between the original and updated container images. Since container images encapsulate applications and their dependencies within layered file systems, they can be unpacked into standard directory hierarchies for analysis.
The corresponding workflow is illustrated in Algorithm~\ref{alg:diff-id-full}.
The process begins with a directory-level delta analysis, which categorizes changes as \textit{added}, \textit{removed}, or \textit{modified} directories (line 1). For each \textit{modified} directory, a recursive file-level delta analysis is triggered, similarly categorizing individual files (line 2). To determine whether a file has been modified, \toolName performs content hash comparisons (line 3). Files with identical names but differing hash values are labeled as \textit{modified} and are subjected to type-specific, fine-grained content comparisons (lines 5-17). File types are broadly classified into textual and non-textual (binary) categories, each requiring a distinct analysis method.

For textual files such as Python scripts, we perform a line-level differential analysis, as illustrated in lines 5-9. Each file is abstracted as an ordered sequence of lines, and the differences between file versions are modeled as transformations from one sequence to another. To compute these transformations, we construct an edit graph where each node represents a position in the original and updated sequences, and each diagonal in the graph captures a possible sequence alignment (line 7). The objective is to identify the shortest edit path that converts the original version into the updated one, minimizing the number of insertions required.
% that models the file as an ordered sequence of lines (lines 5-9). According to ordered sequences of the same file from two application versions, our analysis constructs the corresponding the edit graph. It aims to find the shortest edit path that transforms one sequence into another by minimizing the number of intentions and deletions. The core is to traverse the edit graph where each node represents a position in the two sequences, and each diagonal in the graph corresponds to a potential sequence alignment. 
To optimize performance for large files, our line-level differential analysis adopts a bidirectional divide-and-conquer variant, which concurrently computes forward and reverse search paths until they converge at the midpoint of the edit graph (line 8). Our analysis outputs a sequence of edit operations, indicating insertions, deletions, and retentions, captured as an edit path over line indices.

For non-textual files, e.g., compiled executables or bytecode, direct byte-wise comparison leads to overly fine granularity and inefficiency in delta generation. Even a single-byte insertion or deletion can shift subsequent byte offsets, leading to widespread, non-semantic changes across the entire file. Instead, we adopt a content-defined chunk-level differential analysis (lines 10-17), which computes rolling hash values over a sliding window to detect stable chunk boundaries based on content patterns. When the computed hash satisfies a predefined condition (e.g., a particular number of trailing zero bits, similar to the Rabin Fingerprint method~\cite{Rabinfingerprint}), a chunk boundary is declared (lines 12-13). Each chunk is individually hashed to produce a signature that uniquely represents its content. Our analysis then compares the hash sequences of chunks to identify changed chunks via the similar edit graph construction and path search (lines 14-15). Meanwhile, these chunks are annotated with their corresponding byte numbers, which serve as indicators of content modifications (line 16).
This analysis is more effective than direct byte-wise comparison, thereby enabling stable segmentation and minimizing unnecessary recomputation. 
% It also localizes the impact of small changes, preventing widespread chunk misalignment. 
It outputs a list of changed content chunks, each specified by its byte offset range.

Our hybrid differential strategy, i.e., line-based for textual content and chunk-based for compiled binaries, balances semantic accuracy with computational efficiency, forming the basis for minimal and bandwidth-efficient updates.

\begin{algorithm}[t]
\caption{Content-Aware Differential Identification}
\label{alg:diff-id-full}
\KwIn{

$I_{orig}$: original container image\\
$I_{upd}$: updated container image
}
\KwOut{

$changed\_dirs$: list of changed directories\\
$changed\_files$: list of changed files\\
$edit\_paths$: edit paths for same textual files\\
$chunk\_diffs$: changed chunks for same binary files\\
}

$changed\_dirs \leftarrow \texttt{CompareDir}(I_{orig}, I_{upd})$\;
$changed\_files \leftarrow \texttt{CompareFile}(I_{orig},I_{upd})$\;

\ForEach{$file$ in $changed\_files$}{
  \If{$\texttt{Hash}(file_{orig}) \neq \texttt{Hash}(file_{upd})$}{
    \If{$\texttt{IsTextual}(file)$}{
      \texttt{// Textual File Analysis}\;
      $edit\_graph \leftarrow \texttt{BuildEditGraph}(file_{orig}, file_{upd})$\;
      $edit\_paths \leftarrow \texttt{BidirectionalEdit}(edit\_graph)$\;
      \texttt{StoreEditPath}($file$, $edit\_paths$)\;
    }
    \Else{
      \texttt{// Binary File Analysis}\;
      $chunks_{orig} \leftarrow \texttt{Chunkify}(file_{orig})$\;
      $chunks_{upd} \leftarrow \texttt{Chunkify}(file_{upd})$\;
      $edit\_graph \leftarrow \texttt{BuildEditGraph}(chunks_{orig}, chunks_{upd})$\;
      $chunk\_diffs \leftarrow \texttt{BidirectionalEdit}(edit\_graph)$\;
      \texttt{ChunkWithByteNum}$(chunk\_diffs)$\;
      \texttt{StoreChunkDiffs}($file$, $chunk\_diffs$)\;
    }
  }
}
\Return $changed\_dirs$, $changed\_files$, $edit\_paths$, $chunk\_diffs$
\end{algorithm}

% \textbf{Complexity Analysis of Algorithm~\ref{alg:diff-id-full}.} Let $F$ denote the number of changed files identified in the image comparison process. For each changed file, let $L$ be the number of lines in a textual file and $C$ be the number of chunks in a binary file. We assume that $F_t$ and $F_b$ represent the number of changed textual and binary files, respectively. The line-level difference analysis uses a bidirectional edit strategy, which has a time complexity of $O(L log L)$ per textual file. For binary files, content-defined chunking takes $O(B)$ time per file, and chunk-level comparison incurs an additional $O(C log C)$, where $B$ is the byte size of the file. Thus, the overall time complexity is $O(F_t * L log L + F_b * (C log C))$. The space complexity of Algorithm~\ref{alg:diff-id-full} is $O(F * max(L, C))$, accounting for the temporary storage of edit graphs, intermediate diffs, and chunk metadata. Since the algorithm incrementally processes files and avoids full in-memory representations of entire images, it scales well with the number of modified files and supports efficient delta extraction.

\noindent \underline{\textit{Complexity Analysis of Algorithm~\ref{alg:diff-id-full}.}} Let $f$ be the number of changed files, with $f_t$ and $f_b$ denoting the number of modified textual and binary files, respectively. For a textual file with $l$ lines, the bidirectional line-level comparison incurs a time complexity of $O(l \log l)$. For binary files, content chunking takes $O(b)$ time, and chunk comparison adds $O(c \log c)$, where $b$ is the file size in bytes and $c$ is the number of chunks. Thus, the total time complexity is $O(f_t * l log l + f_b * c log c)$. The space complexity is $O(f * \max(l, c))$, accounting for edit graphs, intermediate diffs, and metadata. As the algorithm processes files incrementally without loading full images into memory, it scales with the file number and supports delta extraction.

\begin{table}[t]
\footnotesize
 \caption{The Application Metadata Representation.}
 % \vspace{-3mm}
    \label{tab:metadata}
    \begin{tabular}{c|c|c}
    \hline
    \textbf{Update Type ($Type$)} &  \textbf{Names ($FilePath$)} & \textbf{Operations ($EditOps$)} \\
    \hline
    
    Type-D & \tabincell{c}{$Directory_{1}$ \\$Directory_{2}$} & \tabincell{c}{I \\ D}   \\ \hline

    Type-F & \tabincell{c}{$File_{1}$ \\ $File_{2}$}  & \tabincell{c}{I \\ D} \\ \hline

    Type-T & $TextualFile_{1}$ & R $l_1$, D $l_2$, I $l_3$, ...   \\ \hline
    
    Type-B & $Non$-$textualFile_{1}$ & R $c_1$, D $c_2$, I $c_3$, ...   \\ \hline

\end{tabular}
\vspace{-3mm}
\end{table}

\textit{\textbf{Step 2: Application Metadata Representation.}} This phase establishes a unified and expressive metadata representation for the upload package, which captures both structural and semantic changes derived from the differential identification phase (Step 1). Specifically, based on the outputs of Step 1, each change is encapsulated in a structural tuple format, enabling standardized processing. As shown in the Table~\ref{tab:metadata}, each metadata entry is expressed as a tuple $\langle Type, FilePath, EditOps\rangle$. $Type$ encodes the category and granularity of the change. It includes: Type-D for added or deleted directories, Type-F for newly added or removed files, Type-T for modified textual files, and Type-B for modified binary files. $FilePath$ denotes the file or directory names, whose file locations are relative to the container image. $EditOps$ encodes the specific edit operations. For Type-D and Type-F, the operations are simplified to atomic Insert (I) and Delete (D) tags. For Type-T, the operations are represented at line-level granularity as an ordered list of instructions: $EditOps$ = $\{$ R $l_1$, D $l_2$, I $l_3$, ...$\}$, where R, D, and I denote line retain, delete, and insert, respectively, and each $l_i$ specifies the number of lines affected. For Type-B, the operations are represented at the chunk level $\{$ R $c_1$, D $c_2$, I $c_3$, ...$\}$, where each $c_i$ denotes the number of chunks involved in the corresponding operation. The size of each chunk, measured at the byte level, is determined during the chunkification process in Step 1. In particular, with the exception of edit-operation metadata, the actual data of inserted code is extracted and stored separately within the upload package using a naming scheme consistent with the original file. This design ensures deterministic reconstruction during the subsequent onboard update process.
To further minimize uplink bandwidth consumption, the upload package is compressed using a standard lossless compression such as Gzip. This exploits the inherent redundancy in source code and textual diffs, reducing the data size and preserving the semantic structure.

\subsection{Onboard Update Component}

\begin{algorithm}[t]
\caption{Onboard Reconstruction for a File}
\label{alg:onboard-update}
\KwIn{

$init\_file$: list of original lines/chunks (from the old file version)\\
$edit\_ops$: sequence of edit operations $(op, len)$\\
% $diff\_list$: index list of updated lines or chunks\\
$seg\_list$: list of added code segments (aligned with insert operations)
}
\KwOut{

$updated\_file$: reconstructed target file version
}

$index\_init \leftarrow 0$ \tcp*{Pointer in $init\_file$}

% $index\_diff \leftarrow 0$ \tcp*{Pointer in $diff\_list$}

$index\_seg \leftarrow 0$ \tcp*{Pointer in $seg\_list$}

\ForEach{$(op, len)$ in $edit\_ops$}{
    \uIf{$op = \texttt{'R'}$}{
        $index\_init \leftarrow index\_init + len$ \tcp*{Retain original content}
    }
    \uElseIf{$op = \texttt{'D'}$}{
        delete $init\_file[index\_init : index\_init + len]$ \tcp*{Remove content}
    }
    \uElseIf{$op = \texttt{'I'}$}{
        $segment \leftarrow seg\_list[index\_seg]$ \;
        insert $segment$ at $init\_file[index\_init]$ \;
        % $index\_diff \leftarrow index\_diff + len$ \tcp*{Advance diff pointer} \;
        $index\_init \leftarrow index\_init + len$ \;
        $index\_seg \leftarrow index\_seg + 1$ \tcp*{Advance to next segment}
    }
}
\Return $init\_file$ as $updated\_file$
\end{algorithm}

% This component takes as input the update package and produces a runnable instance of the new application version as output. This component comprises two stages: Onboard Application Reconstruction and Fault-tolerant Recovery.

This component transforms the original application image into its updated version directly on the satellite, leveraging the structured metadata received in the upload package. It comprises Onboard Application Reconstruction and Fault-Tolerant Recovery, as shown in Step 3 and Step 4 of Fig.~\ref{fig:overview}.

\textit{\textbf{Step 3: Onboard Application Reconstruction}}: During this phase, \toolName processes the original application version by interpreting the associated edit operations and performing a fine-grained application reconstruction. For added or deleted directories and files, \toolName directly applies the corresponding addition or deletion operations. For each file to be updated, \toolName iteratively applies edit instructions, i.e., R (retain), D (delete), and I (insert), over lines or chunks. Insertions are performed using pre-extracted code segments, with file-level consistency preserved via unique identifiers and alignment metadata. Algorithm~\ref{alg:onboard-update} outlines the onboard reconstruction logic for the updated file. The procedure consumes three inputs: (1) the original content of the file ($init\_file$), (2) the edit operations ($edit\_ops$), and (3) the delta code segment list ($seg\_list$) containing the newly inserted content. It outputs the reconstructed target file ($updated\_file$). First, this algorithm uses two pointers to maintain and traverse the content of the original file and delta code segment list (lines 1–2). For each edit operation, the algorithm advances the file pointer in case of R (lines 4-5), deletes a block of content for D (lines 6-7), and inserts a new segment from $seg\_list$ at the current position for I (shown in lines 8-12).

% \toolName parses the update package of metadata representation to incrementally transform the original application files into their updated counterparts. \toolName iteratively applies a sequence of edit operation instructions, such as R, D, or I lines (for text files) or chunks (for binary files), to accurately reconstruct the target version. Particularly, insertions are performed using content retrieved from pre-extracted delta code segments, with file-specific alignment ensured via unique file identifiers. This line-wise (or chunk-wise) reconstruction of edit operations enables deterministic and accurate updates. 

% Algorithm~\ref{alg:onboard-update} presents the onboard reconstruction procedure for a file. It is designed to transform an outdated version of a file into its updated version by sequentially applying a precomputed sequence of edit operations. The input consists of three elements: the original file content ($init\_file$), the added code segments $seg\_list$, and the $edit\_ops$, which encodes the transformation using a sequence of symbolic operations of the form $(op, len)$, where $op \in {\texttt{R, D, I}}$. Algorithm~\ref{alg:onboard-update} maintains two indices to track the current positions in the original file and the code segment list, respectively (lines 1-2). For each operation in the edit path, R advances the pointer in the original file without modification, D removes a specified number of lines or chunks from the current position, and I adds new content from the segment list at the current location in the original file (lines 4-12).

\noindent \underline{\textit{Complexity Analysis of Algorithm~\ref{alg:onboard-update}.}} Let $n$ denote the number of edit operations, $m$ the number of lines/chunks in the original file, and $k$ the number of inserted segments. The time complexity is $O(n+k)$, since each operation is processed sequentially, and insertions require copying new segments. Retain and deletion operations incur constant-time pointer adjustments. The space complexity is $O(m+k)$, as memory is needed to store the original file and the delta segments.

\textit{\textbf{Step 4: Fault-Tolerant Recovery.}} This phase incorporates a fault-tolerant recovery mechanism to ensure application availability in the presence of failures. This mechanism automatically detects and responds to failures that may arise either during the update process or post-update execution. 
% \wjf{Failures occurring during the update process are captured through non-zero exit codes, serving as indicators of unsuccessful updates. In contrast, failures during post-update execution are detected via an event-triggered detection approach through the container runtime interface, which continuously monitors the container's status. Abnormal terminations are identified when the associated exit code is non-zero, enabling timely detection of runtime failures.} 
Failures during the update process are indicated by non-zero exit codes, which mark unsuccessful updates. Post-update execution failures are detected by an event-triggered mechanism in the container runtime interface that continuously monitors container status. Abnormal terminations with the associated non-zero exit codes enable timely identification of runtime failures.
Upon detecting failure events, \toolName triggers an immediate rollback procedure without requiring remote assistance from the ground station.

Our recovery strategy leverages the native layered architecture of containers, in which each application version is encapsulated as a discrete image layer. This structure enables version tracking and facilitates instant restoration. As illustrated in Fig.~\ref{fig:recovery}, when an application update (e.g., image layer V1.1) is identified as faulty, \toolName rollbacks to the last stable image layer (e.g., V1.0) by replacing the uppermost layer. The rollback is automatically performed and with minimal latency, eliminating the need for re-downloading or reinstallation from the ground station.
Compared to image, file, or patch recovery strategies, discussed in experimental evaluation in Section~\ref{sec:rq3faulttolerant}, our layer rollback strategy offers a balanced trade-off between recovery overhead and storage overhead.

\begin{figure}[t]
	\centering
    \includegraphics[width=0.47\textwidth]{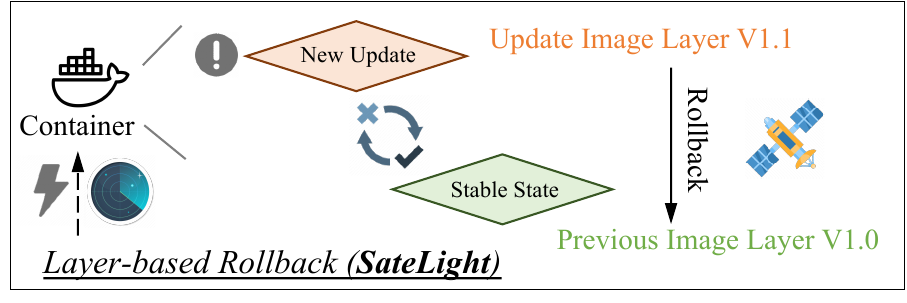}
    % \vspace{-3mm}
    \caption{Fault-tolerant recovery mechanism of \toolName.}
    \label{fig:recovery}
    \vspace{-3mm}
\end{figure}

% it is to ensure the reliability of application updates to avoid potential in-flight failures or data corruption. \toolName incorporated a recovery mechanism, which is designed to detect and respond to failures that may occur during or after the update process. Upon identifying an anomaly, \toolName initiates an automated rollback procedure. It leverages the inherent layered structure of container images, which enables efficient version tracking and reversion. Rather than re-downloading or reapplying the original application, the system reverts the container to a previously committed base layer representing the last known stable version. This layered snapshot-based strategy ensures the atomicity of update transactions and significantly reduces recovery latency. In addition, compared to the fine-grained content-level recovery way, this exhibits the lowest overhead, as demonstrated by our supplementary experimental results \wjf{(see Section~XX)}.

% Add Fault-Tolerant Recovery

\subsection{Prototype Implementation} 
We implement \toolName as a prototype in Python. The design principles of \toolName are language-agnostic and applicable to satellite applications across different programming languages. It is built upon the foundational ideas, including dynamic programming, divide-and-conquer strategy, hash computation, and lossless compression, which are extended and adapted to support the analysis requirements of \toolName.

% Our implementation builds upon and extends the capabilities provided by the dynamic programming, hash calculation, and gzip compression functionalities, which serve as the foundation for the core analysis components.

% \toolName is implemented as a Python prototype. Note that the design principles of \toolName behind different languages are general. In our approach, the main analysis builds on the foundation and improvements of XX, XX, XX, and XX libraries.

\section{Experimental Evaluation}\label{sec:evaluation}

We first present our research questions, followed by the baselines used for comparison and the satellite application dataset. We then describe the experimental settings.

\subsection{Research Questions}

% \textit{\textbf{RQ1: To what extent can \toolName reduce the size of satellite application updates, compared to baselines?}} This RQ investigates the effectiveness of \toolName in minimizing the amount of upload data required for application updates by leveraging differential and content-aware analysis, which is crucial for reducing transmission costs in bandwidth-constrained satellite environments.

% \textit{\textbf{RQ1: How much can \toolName improve transmission performance during application uploads from the ground to the satellite, compared to baselines?}} This RQ examines the impact of \toolName on the transmission latency, measuring how its content-aware difference identification strategy accelerates the update delivery process.

\textit{\textbf{RQ1 (Upload Transmission Overhead):}} \textit{How much can \toolName reduce the upload transmission overhead of satellite applications compared to baselines?} We assess the effectiveness of \toolName's content-aware upload strategy through data size and transmission latency comparisons with baselines.

\textit{\textbf{RQ2 (Onboard Update Overhead):}} \textit{How is the onboard update overhead of \toolName relative to baselines?} We measure the onboard update latency introduced by the application reconstruction strategy of \toolName, relative to baselines.

% This RQ examines the onboard update latency incurred by \toolName’s onboard update mechanism when processing and applying fine-grained content modifications to the original application. 

% \textit{\textbf{RQ3: How is fault tolerance of \toolName?}} This RQ evaluates the effectiveness of fault tolerance of \toolName by measuring the time cost involved in restoring application functionality after update failures.

\textit{\textbf{RQ3 (Recovery Overhead):}} \textit{How effective is the fault-tolerant recovery mechanism of \toolName during update failures?} We assess the recovery performance overhead of \toolName's layer-based rollback design under update failures.

\textit{\textbf{RQ4 (Real-World Case):}} \textit{How practical is \toolName for real-world in-orbit satellite deployment?} We validate the onboard update process of \toolName via a case study, involving their deployment and execution on an in-orbit satellite.

% \textit{\textbf{RQ4 (Real-World Case):}} \textit{How practical is \toolName for real-world on-orbit satellite deployment?} We validate the onboard update process of \toolName through a case study, involving their deployment and execution on an operational on-orbit satellite.

% via a case study, which is deployed on a real-world on-orbit satellite.

% This RQ aims to validate the practicality of the onboard reconstruction application based on the uploaded update package via a real-world on-orbit satellite.

% \textit{\textbf{RQ5: How does \toolName perform compared to state-of-the-art methods?}}
% This RQ compares \toolName with other update baselines in terms of update size, transmission latency, and update latency.

\subsection{Compared Baselines}

To the best of our knowledge, no existing application update approaches have specifically addressed the challenges of satellite computing. To evaluate \toolName, we compare it against the following potential baselines in this scenario:

$\bullet$ \textbf{\textit{Baseline 1 (Image-level approach):}} The entire container image including the updated application version is uploaded and deployed on the satellite. This baseline performs full image transmission without any onboard reuse or processing.

$\bullet$ \textbf{\textit{Baseline 2 (Application-level approach):}} Only the application layer within the container image that contains the updated application version is uploaded. This approach reuses unchanged layers, and the new image is reconstructed onboard with the updated layer.

$\bullet$ \textbf{\textit{Baseline 3 (File-level approach):}} Within the container image, changed files related to the application are identified. Only these changed files from the new version are collected, compressed, uploaded, and used to replace the original version onboard. This approach avoids the upload of unchanged files.

\subsection{Satellite Application Dataset}

\begin{table}[t]
\footnotesize
 \caption{The representative satellite applications used in this paper.}
 % \vspace{-3mm}
    \label{tab:testapplication}
    \begin{tabular}{p{0.7cm}|p{2.2cm}|p{3.2cm}|p{1.1cm}}
    \hline
    \textbf{AppID} & \textbf{Applications} &  \textbf{Task description} & \textbf{Languages} \\
    \hline
    App1& Object Detection~\cite{xing2024deciphering} & It detects the specific objects in an image. & Python  \\
    \hline
    App2& Core Network~\cite{xing2023earth} & It includes network functions for AMF, SMF, and UPF. & C  \\
    \hline
    App3& Image Encoding~\cite{xing2024deciphering} & It transforms and encodes images. & Python  \\
    \hline
    App4& Client Cache~\cite{pfandzelter2024komet} & It implements a read-write cache. & Python, Go  \\
    \hline
    App5& Multi-Stage Image Compression~\cite{zhai2024seco} & It implements a five-stage image compression. & MATLAB  \\
    \hline
    App6& Ship Detection~\cite{liu2024orbit} & It detects ships from satellite images. & Python  \\
    \hline
    App7& Tracking Algorithm~\cite{tsog2018intelligent} & It implements a feature tracking task. & C++\\
    \hline
    App8& Data Compression~\cite{ouyang2023joint} & It reduces data volume via neural-based autoencoders. & Python\\
    \hline
    App9& Attitude Determination~\cite{attitudedetermination} & It estimates the satellite’s attitude. & JavaScript\\
    \hline
    App10& Change Detection and Encoding~\cite{du2025earth} & It lightly detects and encodes changes. & C, C++\\
    \hline
\end{tabular}
\vspace{-4mm}
\end{table}

We collect 10 representative satellite applications mainly from the published relevant papers, encompassing tasks such as object detection~\cite{xing2024deciphering}, core network functionality~\cite{xing2023earth}, and attitude determination~\cite{attitudedetermination}. Their detailed descriptions are provided in Table~\ref{tab:testapplication}. The implementations span a range of commonly used satellite-related programming languages, including Python, C, Go, C++, Matlab, and JavaScript. All applications are designed to be stateless, meaning that previous computations and results do not influence the current processing. This design is both reasonable and well-suited for handling raw data generated by satellites~\cite{zhang2024resource}.

Our software update scope encompasses functional and non-functional changes to satellite application code. Due to the absence of satellite application update datasets, we construct synthetic variants of representative applications exhibiting varying degrees of code modification.
In practice, software evolves through version updates that introduce new features or address defects. Empirical studies have quantified the extent of such changes across diverse systems. These studies have shown that version updates generally affect less than approximately 10\% or 20\% of source components~\cite{codepercentage, mcintosh2016empirical, koch2007software}, while more substantial releases may modify more of the codebase with upper bounds generally remaining below 50\%~\cite{neamtiu2013towards}. Grounded in these empirical findings, we model application updates using approximately 10\% and 20\% of code modifications relative to the original application, respectively, and simulate large-scale changes with a modification level of around 50\%.
% Grounded in these empirical results, we adopt about 10\% and 20\% of contextual changes relative to the original application to represent minor and moderate updates, respectively, and use about 50\% modification to simulate large-scale changes.

To isolate update effectiveness from application-level bugs, all variants are \textit{task-goal intent equivalent}, preserving high-level task objectives while varying internal implementations. Specifically, we first determine the change ratio, compute the modification size, and then randomly select files and code regions for manual modification. Code updates involve insertions and deletions. Insertions include semantically neutral constructs such as comments, logging statements, inactive conditionals, and unused variables. Deletions target non-functional elements like redundant comments, unused imports, and logs. For compiled languages, comment modifications are excluded from consideration, as they are eliminated during preprocessing and have no effect on the binary. This generation process ensures that the modified applications remain task-goal intent equivalent to their original versions and execute without errors. We validate the executability of these variants through real execution.

% These update levels are constructed manually through a controlled process described as follows. First, we determine the desired percentage of code change and randomly select the corresponding files and code regions to be modified. To ensure the application execution, we adopt a \textit{functionality-preserving} principle throughout the modification process. Code updates are performed using insertion and removal operations. Specifically, insertions consist of behaviorally neutral code such as comments, logging statements, inactive conditionals, and unused variable declarations. Removals target non-functional elements including comments, unused imports, and logging statements. For the compiled languages, such as C language, comments are generally discarded during the prepossessing phase and do not influence the compiled output. As such, we do not consider modifications to comments. Overall, these strategies ensure that the modified applications remain functionally equivalent to their original versions and execute without errors, thereby creating a reliable evaluation for update-aware analysis.

To measure the extent of code modifications between two versions of a satellite application, we present a quantitative metric. This metric captures the proportion of the new version's content that differs from the previous version, thereby offering an interpretable indication of update magnitude. It is formally defined as $1- \frac{S_{preserved}}{S_{upd}} = 1 - \frac{\sum_{i=1}^{N} len_{byte}(u_i)}{S_{upd}}$.
% \begin{align}
% \footnotesize
%     DR = 1- \frac{S_{preserved}}{S_{upd}} = 1 - \frac{\sum_{i=1}^{N} len_{byte}(u_i)}{S_{upd}}
%     \label{eq:precision}
% \end{align}
$S_{preserved}$ denotes the total size, in bytes, of code segments that are preserved across both the old and new versions of the application. These segments are identified by detecting common subsequences through our differential analyses, which support granularity at both the line and chunk levels of source code. The size is computed as the sum of the byte lengths of the preserved units $u_i$, with each unit's size given by $len_{byte}(u_i)$. $S_{upd}$ denotes the total size in bytes of the updated version of the application. By computing the relative complement of the retained content within the new version, it effectively captures the extent of code modifications.

\subsection{Experimental Settings}

\noindent \textbf{Experimental Environment.} 
% To support onboard sensing, computation, and control tasks, the aerospace field adopts Commercial Off-the-shelf (COTS) computing devices on the satellite. Raspberry Pi 4B industrial module~\cite{zhang2024resource} as a representative COTS devices, is widely adopted in satellite research~\cite{zhang2024resource, chen2024energy, xing2024deciphering} and has been successfully deployed in multiple in-orbit missions~\cite{xing2024deciphering, Pidata}, demonstrating its feasibility and stability.
To support onboard sensing, computation, and control tasks, the aerospace field increasingly adopts Commercial Off-the-Shelf (COTS) computing devices in satellite systems~\cite{xing2024deciphering, chen2024energy}. The Raspberry Pi 4B industrial module~\cite{zhang2024resource}, as a representative COTS computing device, has been widely used in satellite research~\cite{zhang2024resource, chen2024energy, xing2024deciphering} and successfully deployed in multiple in-orbit missions~\cite{xing2024deciphering, Pidata}, demonstrating its feasibility and operational stability.
% the satellite is equipped with a Raspberry Pi 4B industrial module~\cite{zhang2024resource}. This hardware module is widely adopted in satellite research~\cite{zhang2024resource, chen2024energy, xing2024deciphering} and has been successfully deployed in multiple in-orbit missions~\cite{xing2024deciphering, Pidata}, demonstrating its feasibility and stability.
Thus, to answer RQ1-RQ3, we use the Raspberry Pi 4B as the satellite simulation environment. On top of the hardware, the evaluated applications are containerized and executed in isolated environments using Docker~\cite{docker}, the industry and open-source standard for container deployment. Prior studies~\cite{wang2023satellite, pfandzelter2024komet} have demonstrated the practicality of deploying containerized applications on satellites. To answer RQ4, we deploy the onboard functionality of \toolName on a real-world satellite (called BUPT-2) equipped with onboard computing capabilities and operating in a Sun-synchronous orbit. Satellite operations are supported by two ground stations: a telemetry, tracking, and control (TT\&C) station, which manages command uplink and satellite control functions; and a data transmission station, responsible for uploading application update packages and downlinking payload data generated onboard.

% a telemetry, tracking, and control station, which handles command uploading and satellite control, and a data transmission station, responsible for uploading the application update data and downlinking payload data generated onboard.

% with onboard computing capabilities. This satellite operates in a Sun-synchronous orbit. We use two ground stations to control the satellite. The telemetry, tracking, and control station enables the loading of command sequences onto the satellite for execution and control. The data transmission station is responsible for receiving the payload data generated by the satellite.

\noindent \textbf{Evaluation Metrics.} We evaluate \toolName and baselines using the collected satellite applications with code modifications (about 10\%, 20\%, and 50\%). Each application has three update variants. Evaluation metrics are as follows:

% Evaluation metrics include transmission latency, onboard update latency, and end-to-end latency. The detailed descriptions are as follows:

$\bullet$ \textit{Transmission latency}: We estimate this latency based on the specific uplink bandwidth from the ground station to the satellite. Consistent with prior studies that commonly assume uplink bandwidth in the range of hundreds of kbps~\cite{tao2023transmitting, devaraj2019planet, vasisht2021l2d2, wang2023satellite}, we adopt a representative value of 200 kbps, as suggested in the work~\cite{vasisht2021l2d2}. The uplink bandwidth is a constant, as the uplink leverages the low-frequency S-band to communicate~\cite{SBand, zhang2024resource}. It is computed by dividing the total upload package size (in bits) by the uplink bandwidth (in bits per second).

 $\bullet$ \textit{Onboard update latency}: This latency is computed as the duration between the initiation of the onboard update process and the completion of the update. It reflects the internal processing time required for onboard application reconstruction.

 $\bullet$ \textit{Correctness}: This metric evaluates whether the application has been updated accurately. It identifies consistency between the onboard updated application and the corresponding version validated for successful execution at the ground station.

% $\bullet$ \textit{End-to-end latency}: The overall latency experienced during the satellite application update process is calculated as the sum of transmission latency and onboard update latency.

\noindent \textbf{Experimental Repetitions.} Each experiment is repeated ten times, and the mean is reported to mitigate the random impact.

\section{Experimental Results}

\subsection{RQ1 Results (Upload Transmission Overhead)}

\begin{table*}[t]
\centering
\caption{RQ1: Transmission Overhead Results (data size and \textbf{transmission latency}) of \toolName and baselines.}
 % \vspace{-3mm}
    \label{tab:RQ1transmission}
\begin{tabular}{p{0.6cm}|p{0.5cm}|r|r|r|r|r|r|r|r|r|r|r}
\begin{tabular}[c]{@{}l@{}}\textbf{ID}\end{tabular}&  \begin{tabular}[c]{@{}l@{}}\textbf{Level}\end{tabular} & \multicolumn{2}{|c}{Baseline 1 (B1)} & \multicolumn{2}{|c}{Baseline 2 (B2)}  & \multicolumn{2}{|c}{Baseline 3 (B3)}  & \multicolumn{2}{|c}{\textbf{\toolName}} & \multicolumn{3}{|c}{\textbf{Improvement of \toolName $\uparrow$}}\\
\hline
   &    & \begin{tabular}[c]{@{}c@{}}Data Size \\ (KB)\end{tabular}  & \begin{tabular}[c]{@{}c@{}}Latency \\ (s)\end{tabular} & \begin{tabular}[c]{@{}c@{}}Data Size \\ (KB)\end{tabular} & \begin{tabular}[c]{@{}c@{}}Latency \\ (s)\end{tabular} & \begin{tabular}[c]{@{}c@{}}Data Size \\ (KB)\end{tabular} & \begin{tabular}[c]{@{}c@{}}Latency \\ (s)\end{tabular} & \begin{tabular}[c]{@{}c@{}}Data Size \\ (KB)\end{tabular} & \begin{tabular}[c]{@{}c@{}}Latency \\ (s)\end{tabular} & \begin{tabular}[c]{@{}c@{}}(vs. B1)\\Latency\end{tabular} & \begin{tabular}[c]{@{}c@{}}(vs. B2)\\Latency\end{tabular} & \begin{tabular}[c]{@{}c@{}}(vs. B3)\\Latency\end{tabular}  \\
\hline
\multirow{3}{*}{App1} & 10\% &942,701.00 & 38,613.03&188,845.50&7,735.11&146.96&6.02&48.64&1.99& 99.99\% &99.97\% & 66.94\%   \\
\cline{2-13}
& 20\% &943,308.00&38,637.90&189,061.38&7,743.95&331.91&13.60&154.34&6.32& 99.98\% &99.92\% &53.53\%  \\
\cline{2-13}
& 50\%   &945,335.00&38,720.92&189,819.63&7,775.01&905.87&37.10&709.26&29.05& 99.92\% & 99.63\% & 21.70\%  \\
% \hline
% & 80\%   &952046.00&38995.80&191111.74&7827.94&2333.12&95.56&2091.67&85.67&99.78\% &98.91\%&10.35\% \\  
\hline
\multirow{3}{*}{App2} & 10\% &127,516.50&5,223.08&11.50&0.47&11.01&0.45&2.24&0.09& 100.00\% &80.85\% &80.00\%  \\
\cline{2-13}
& 20\% &127,517.00&5,224.10&11.69&0.48&11.01&0.45&2.24&0.09&100.00\%& 81.25\%& 80.00\% \\
\cline{2-13}
& 50\%   &127,516.50&5,223.08&11.49&0.47&10.82&0.44&3.65&0.15&100.00\%& 68.09\%&65.91\% \\
% \cline{2-13}
% & 80\%   &127538.50&5223.98&32.91&1.35&10.82&0.44&5.98&0.25&99.99\%&81.82\%&44.70\% \\  
\hline
\multirow{3}{*}{App3} & 10\% &62,805.50&2,572.51&0.88&0.04&0.67&0.03&0.29&0.01&100.00\% &75.00\% &66.67\%  \\
\cline{2-13}
& 20\%   &62,805.50&2,572.51&0.91&0.04&0.69&0.03&0.34&0.01&100.00\%& 75.00\%&66.67\%  \\
\cline{2-13}
& 50\%   &62,805.50&2,572.51&1.00&0.04&0.77&0.03&0.45&0.02&100.00\%&50.00\% &33.33\% \\  
\hline
\multirow{3}{*}{App4} & 10\% &386,661.50&15,837.66&9.77&0.40&8.71&0.36&1.47&0.06&100.00\%&85.00\%&83.33\%  \\
\cline{2-13}
& 20\%   &386,664.00&15,837.76&11.78&0.48&9.79&0.40&1.96&0.08&100.00\%&83.33\%&80.00\%  \\
\cline{2-13}
& 50\%   &386,665.00&15,837.80&12.83&0.53&13.03&0.53&3.08&0.13&100.00\%&75.47\%&75.47\% \\  
\hline
\multirow{3}{*}{App5} & 10\% &2,310,128.50&94,622.86&11.51&0.47&9.97&0.41&1.93&0.08 &100.00\% &82.98\%&80.49\% \\
\cline{2-13}
& 20\%  &2,310,130.50&94,622.95&13.44&0.55&4.46&0.18&3.68&0.15&100.00\%&72.73\%&16.67\% \\
\cline{2-13}
& 50\%   &2,310,141.00&94,623.38&23.79&0.97&22.62&0.93&13.38&0.55&100.00\%&43.30\%&40.86\% \\
\hline
\multirow{3}{*}{App6} & 10\% &743,251.50&30,443.58&205.08&8.40&77.76&3.19&32.41&1.33 & 100.00\%& 84.17\%&58.31\%  \\
\cline{2-13}
& 20\% &743,294.00&30,445.32&247.67&10.14&145.92&5.98&72.12&2.95&99.99\%&70.91\%&50.67\%  \\
\cline{2-13}
& 50\%   &743,540.00&30,455.40&493.39&20.21&412.86&16.91&316.20&12.95&99.96\%& 35.92\%&23.42\% \\  
\hline
\multirow{3}{*}{App7} & 10\% &392,131.00&16,061.69&8.61&0.76&16.50&0.68&2.26&0.09& 100.00\%& 88.16\%&86.76\%  \\
\cline{2-13}
& 20\%   &392,131.00&16,061.69&18.61&0.76&16.52&0.68&4.41&0.18&100.00\%&76.32\%&73.53\%  \\
\cline{2-13}
& 50\%   &392,131.00&16,061.69&18.58&0.76&16.52&0.68&1.35&0.06& 100.00\%& 92.11\%&91.18\% \\  
\hline
\multirow{3}{*}{App8} & 10\% &265,721.00&10,883.93&2.77&0.11&2.29&0.09&0.98&0.04&100.00\%&63.64\% &55.56\%  \\
\cline{2-13}
& 20\%   &265,721.50&10,883.95&3.16&0.13&2.68&0.11&1.34&0.05&100.00\%&61.54\% &54.55\%  \\
\cline{2-13}
& 50\%   &265,723.50&10,884.03&4.90&0.20&4.65&0.19&3.19&0.13&100.00\%&35.00\%&31.58\% \\  
\hline
\multirow{3}{*}{App9} & 10\% &54,557.50&2,234.68&17.76&0.73&4.13&0.17&2.99&0.12&99.99\%&83.56\%&29.41\%\\
\cline{2-13}
& 20\%   &54,557.50&2,234.68&17.75&0.72&6.85&0.28&5.57&0.23&99.99\%&68.06\%&17.86\%\\
\cline{2-13}
& 50\%   &54,575.00&2,235.39&35.14&1.44&21.01&0.86&20.22&0.83& 99.96\%& 42.36\%& 3.49\%\\  
\hline
\multirow{3}{*}{App10} & 10\% &489,989.00&20,069.95&28.95&1.19&26.12&1.07&3.05&0.12 & 100.00\%& 89.92\%&88.79\%  \\
\cline{2-13}
& 20\%   &489,989.00&20,069.95&29.04&1.19&26.19&1.07&4.92&0.20&100.00\%& 83.19\%&81.31\%  \\
\cline{2-13}
& 50\%   &489,989.00&20,069.95&28.98&1.19&26.13& 1.07&16.01&0.66& 100.00\%&44.54\% &38.32\% \\  
\hline
% \textbf{Max} & & & & & & & & & & 100.00\% & 99.97\% & 91.18\%\\
% \hline
\textbf{Mean} & & & & & & & & & & \textbf{99.99\%} & \textbf{73.06\%} & \textbf{56.54\%}\\
\hline
\end{tabular}
 \vspace{-3mm}
\end{table*}

% To explore the uplink transmission overhead of \toolName, we compare the upload date size and transmission latency compared to three baselines in satellite applications under different levels of code modifications. The specific results are shown in Table~\ref{tab:RQ1transmission}. Results show that \toolName can reduce the upload data size by up to XX\% and transmission latency by up to XX\%, compared to three baselines. Specifically, baseline 1 uploads the data size between XX and XX. On average, its data size is XX. Particularly, for the different variants of the same application, baseline 1 keeps comparable uploaded data size. This is because this baseline uploads the entire container image, which is not sensitive to application content. 

To evaluate the transmission overhead of \toolName, we compare it against three baselines across satellite applications, each with three levels of code modifications. Table~\ref{tab:RQ1transmission} reports the data size and transmission latency. Results show that \toolName consistently outperforms all baselines in both metrics. It achieves transmission latency reduction of up to 100.00\%, 99.97\%, and 91.18\% compared to baselines 1, 2, and 3, respectively. On average, \toolName reduces transmission latency by 99.99\%, 73.06\%, and 56.64\%, showing the effectiveness across varied update granularities and application types.

We further analyze the causes of the poor performance observed in the baselines. Baseline 1 incurs the highest overhead, as it uploads the entire container image for every update, irrespective of the actual code change. This results in near-constant data sizes (e.g., about 127,000 KB across all update levels for App2) and high transmission latency (up to 5,224.10 seconds). Baseline 2, which uploads only the application layer, reduces some redundancy by avoiding full-image transfer. However, it still sends the entire layer regardless of how minor the internal code edits are. For instance, in App1 (20\%), it transmits 189,061.38 KB, a reduction from 943,308.00 KB of baseline 1, resulting in latency saving (7,743.95 vs. 38,637.90 seconds). Baseline 3 improves further by uploading only changed files. It performs well on small updates, for example, in App1 (10\%), it transmits 146.96 KB, with a latency of 6.02 seconds. However, its file-level granularity leads to performance degradation as code changes increase. In App1 (50\%), it transmits 905.87 KB, which is more than \toolName. This is due to its lack of semantic awareness, causing even minor changes to trigger whole-file transfers. In contrast, \toolName employs a fine-grained, context-aware differential update, which isolates and transmits only semantically meaningful changes. This results in consistently minimal data transfer across all settings. For instance, in App1 (10\%), it uploads only 48.64 KB, achieving 99.97\% latency reduction over baseline 2, and 66.94\% over baseline 3. In App1 (20\%), despite increased update scope, it keeps latency at 6.32 seconds, still outperforming all baselines. Even in large updates, e.g., App1 (50\%), the latency of \toolName is 29.05 seconds, better than 37.10 seconds of baseline 3. 

We also analyze the mean transmission latency reductions achieved by \toolName under varying levels of code modification. As illustrated in Fig.~\ref{fig:transmission}, \toolName consistently outperforms all three baselines across all levels, with particularly significant improvements under small-scale updates. At the 10\% update level, \toolName achieves mean transmission latency reductions of 100.00\%, 83.32\%, and 69.63\% over baselines 1, 2, and 3, respectively. These reductions highlight the effectiveness of \toolName's fine-grained, content-aware approach, which minimizes unnecessary data transmission by identifying code differences. As the update scope increases, the performance gain gradually decreases, which is expected due to the growing amount of actual content that must be transmitted. Nonetheless, \toolName still maintains a considerable advantage. At the 50\% modification level, it achieves mean reductions of 99.98\% (vs. baseline 1), 58.64\% (vs. baseline 2), and 42.53\% (vs. baseline 3). 
% These findings are consistent with the maximum reduction values shown in Table~\ref{tab:RQ1transmission}, where \toolName achieves up to 100.00\%, 99.63\%, and 91.18\% reductions over the respective baselines at the 50\% modification level. 
This trend also reveals an insight. Baseline 1 is highly sensitive to any update, due to its full-container upload strategy, making the advantage of \toolName nearly absolute regardless of update size. Baseline 2 shows moderate resilience at smaller update levels, but suffers from fixed-layer limitations. Baseline 3 benefits from file-level granularity, yet fails to capture similarities in larger changes, leading to degraded performance under heavier updates.

\begin{figure}[t]
	\centering
    \includegraphics[width=0.45\textwidth]{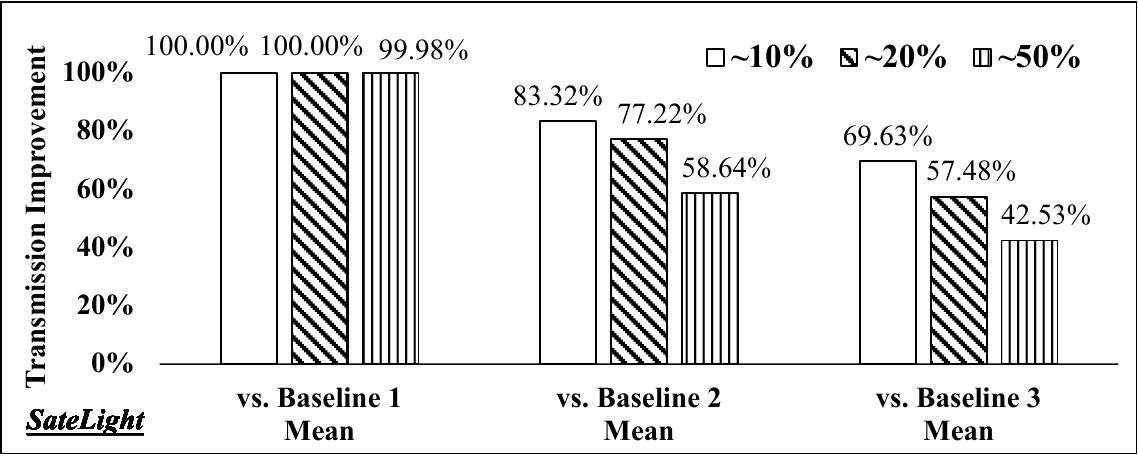}
    % \vspace{-3mm}
    \caption{Mean transmission performance improvement of \toolName under varying code modification levels.}
    \label{fig:transmission}
    \vspace{-4mm}
\end{figure}

Overall, these results confirm that \toolName provides consistent transmission efficiency. This makes it suitable for bandwidth-constrained satellite computing, where reducing uplink time is crucial for timely application updates.

\vspace{-2mm}
\finding{
% \toolName significantly reduces upload transmission overhead, achieving up to XX\% reduction in transmission latency. On average, it outperforms baselines 1, 2, and 3 by XX\%, XX\%, and XX\% in transmission latency, respectively, demonstrating its effectiveness in bandwidth-constrained satellite computing.
\toolName significantly reduces upload transmission overhead. On average, it outperforms three baselines by 99.99\%, 73.06\%, and 56.54\%, respectively. The most pronounced improvements occur at lower code modification levels, with at least 69.63\% mean reductions. These results underscore the effectiveness of \toolName in bandwidth-constrained satellite computing.
}
\vspace{-6mm}

\subsection{RQ2 Results (Onboard Update Overhead)}

\begin{table*}[t]
\centering
\caption{RQ2: Onboard Update Overhead Results (\textbf{onboard update latency}) of \toolName and baselines.}
 % \vspace{-3mm}
    \label{tab:RQ2onboard}
\begin{tabular}{p{0.6cm}|p{0.5cm}|c|r|r|r|c|c|r|r}
\begin{tabular}[c]{@{}c@{}}\textbf{ID}\end{tabular}& \begin{tabular}[c]{@{}l@{}}\textbf{Level}\end{tabular} & \multicolumn{1}{|c|}{Baseline 1} & \multicolumn{1}{|c|}{Baseline 2}  & \multicolumn{1}{|c|}{Baseline 3}  & \multicolumn{1}{|c}{\textbf{\toolName}} & \multicolumn{1}{|c}{\textbf{\toolName}} & \multicolumn{3}{|c}{\textbf{Difference of \toolName}} \\
\hline
   &    & \begin{tabular}[c]{@{}c@{}}Onboard Update \\Latency\end{tabular}  & \begin{tabular}[c]{@{}c@{}}Onboard Update \\Latency (second)\end{tabular}& \begin{tabular}[c]{@{}c@{}}Onboard Update \\Latency (second)\end{tabular} & \begin{tabular}[c]{@{}c@{}}Onboard Update \\Latency (second)\end{tabular} &\multicolumn{1}{|c|}{Correctness}  & \begin{tabular}[c]{@{}c@{}}(vs. B1) \\ Latency\end{tabular} & \begin{tabular}[c]{@{}c@{}}(vs. B2) \\ Latency\end{tabular} & \begin{tabular}[c]{@{}c@{}}(vs. B3) \\ Latency\end{tabular}  \\
\hline
\multirow{3}{*}{App1} & 10\% & --  &32.26&43.42&44.55 & \checkmark & --  & + 12.29 & + 1.13 \\
\cline{2-10}
& 20\% & -- &34.06&46.03&46.47& \checkmark &  -- & + 12.41 & + 0.44 \\
\cline{2-10}
& 50\% & --  &34.92&54.29&58.54& \checkmark & -- & + 23.62 & + 4.25 \\
% \cline{2-12}
% & 80\% & X  &37.59&92.62&119.40&  X &+81.82&+26.78&99.47\%&97.39\%&-8.97\%\\
\hline
\multirow{3}{*}{App2} & 10\% & --  &1.64&1.69&1.74& \checkmark & --  & + 0.10 & + 0.05  
\\
\cline{2-10}
& 20\% & --  &1.64&1.68&1.74&\checkmark & --  & + 0.10 & + 0.06\\
\cline{2-10}
& 50\% & --  &1.64&1.68&1.90& \checkmark & -- & + 0.26 & + 0.22 \\
% \cline{2-12}
% & 80\% & X  & 1.63&1.68&1.93&  X &+0.39&+0.25&99.96\%&27.00\%&-2.35\%\\
\hline
\multirow{3}{*}{App3} & 10\% & --  &0.86&0.90&0.89& \checkmark & --  &+ 0.03 & - 0.01  \\
\cline{2-10}
& 20\% & --  &0.86&0.90&0.90& \checkmark & -- & + 0.04 & 0.00 \\
\cline{2-10}
& 50\% & --  &0.86&0.90&0.90& \checkmark &  -- & + 0.04 & 0.00 \\
\hline
\multirow{3}{*}{App4} & 10\% & --  &5.13&5.18&5.17& \checkmark & -- & + 0.04 & - 0.01  \\
\cline{2-10}
& 20\% & --  &5.11&5.16&5.17&\checkmark & --  & + 0.06 & + 0.01 \\
\cline{2-10}
& 50\% & --  &5.21&5.19&5.32& \checkmark & -- & + 0.11 &+ 0.13\\
\hline
\multirow{3}{*}{App5} & 10\% & --  &120.24&129.34&136.67& \checkmark & -- & + 16.43 & + 7.33\\
\cline{2-10}
& 20\% & --  &121.79&117.15&128.23 & \checkmark &  -- & + 6.44 & + 11.08 \\
\cline{2-10}
& 50\% & --  & 122.74 &116.02&128.41  & \checkmark & -- & + 5.67 & + 12.39 \\
\hline
\multirow{3}{*}{App6} & 10\% & --  &16.65&17.39&17.41& \checkmark & --  & + 0.76 & + 0.02 \\
\cline{2-10}
& 20\% & --  &16.60&16.66&16.86& \checkmark & --  & + 0.26 & + 0.20 \\
\cline{2-10}
& 50\% & --  &16.07&16.25&16.65& \checkmark &  -- & + 0.58 & + 0.40  \\
\hline
\multirow{3}{*}{App7} & 10\% & --  &4.95&5.04&5.10& \checkmark & -- & + 0.15 & + 0.06  \\
\cline{2-10}
& 20\% & --  &5.02&5.12&5.11& \checkmark &--  & + 0.09 & - 0.01 \\
\cline{2-10}
& 50\% & --  &5.03&5.07&5.11& \checkmark & --  & + 0.08 & + 0.04 \\
\hline
\multirow{3}{*}{App8} & 10\% & --  &3.40&3.45&3.46& \checkmark &--  & + 0.06 & + 0.01  \\
\cline{2-10}
& 20\% & --  &3.44&3.47&3.48& \checkmark & -- & + 0.04 & 0.00  \\
\cline{2-10}
& 50\% & --  &3.47&3.51&3.51& \checkmark & -- & + 0.04 & 0.00\\
\hline
\multirow{3}{*}{App9} & 10\% & --  &0.76&0.81&0.81& \checkmark & -- & + 0.05 & 0.00  \\
\cline{2-10}
& 20\% & --  &0.81&0.82&0.82& \checkmark & --  & + 0.01 &0.00\\
\cline{2-10}
& 50\% & --  &0.77&0.82&0.82& \checkmark & -- & + 0.05 & 0.00  \\
\hline
\multirow{3}{*}{App10} & 10\% & --  &6.36&6.29&6.30& \checkmark & -- & - 0.06 & + 0.01  \\
\cline{2-10}
& 20\% & --  &6.45&6.49&6.48& \checkmark & -- & + 0.03 & - 0.01 \\
\cline{2-10}
& 50\% & --  &6.37&6.49& 6.82&  \checkmark & -- & + 0.45 & + 0.33 \\
\hline
% \textbf{Max} &   & X &    &   &   & & X & + 23.62 & + 12.39 \\
% \hline
\textbf{Mean} &   &  &    &   &   & & & \textbf{+ 2.67} & \textbf{+ 1.27}\\
\hline
\end{tabular}
\vspace{-4mm}
\end{table*}

To assess the onboard update overhead introduced by \toolName, we compare its onboard update latency against three baseline strategies. The results, presented in Table~\ref{tab:RQ2onboard}, include both the absolute update latency and the difference relative to each baseline. Note that entries marked as ``--'' indicate no onboard update processing is required, as in baseline 1, which directly executes the full uploaded container image. Despite employing a fine-grained, content-aware update strategy, \toolName incurs only minimal onboard overhead, comparable to or slightly higher than baselines 2 and 3, which rely on coarser update mechanisms. On average, \toolName increases onboard update latency by 2.67 seconds relative to baseline 2 and 1.27 seconds compared to baseline 3. This marginal overhead confirms the efficiency of the onboard reconstruction mechanism of \toolName. For example, in App6 (10\%), the latency of \toolName is 17.41 seconds, while baselines 2 and 3 report 16.65 seconds and 17.39 seconds, respectively, indicating a negligible difference. Even in high-latency scenarios such as App5 (10\%), \toolName incurs 136.67 seconds, which is within 16.43 seconds of baseline 2 and 7.33 seconds of baseline 3.

% We also observe that update latency may not increase with higher code modification levels for both baseline 3 and \toolName, although these approaches reconstruct the application based on the extent of code changes. 
% In contrast, baseline 2 does not exhibit this behavior, since any modification triggers a full application-layer update, resulting in relatively consistent transmission latency, as shown in Table~\ref{tab:RQ1transmission}.

% Our onboard update strategy effectively helps to address the challenge of limited communication bandwidth by ensuring fine-grained update ability onboard. It can achieve 100\% update correctness, as the onboard updated applications are fully consistent with the application update versions validated for successful execution at the ground station. Despite employing fine-grained application construction, \toolName incurs only minimal overhead while ensuring the correctness of updated content.

% Our update strategy mitigates the challenge of limited communication bandwidth by enabling fine-grained update capability directly on the satellite. It achieves 100\% correctness, as shown in Table~\ref{tab:RQ1transmission}, with onboard updated applications fully consistent with the versions validated for successful execution at the ground station. Despite the use of fine-grained application construction, \toolName introduces only minimal overhead while preserving the integrity of the updated content.

Particularly, \toolName achieves 100\% update correctness across all cases, as indicated in the ``Correctness'' column. This ensures that the reconstructed onboard applications are equivalent to their ground-tested counterparts, meeting a fundamental requirement for execution. The use of fine-grained application construction of \toolName introduces only minimal overhead while preserving the integrity of the updated content.

\vspace{-2mm}
\finding{
% \toolName introduces comparable onboard update overhead compared to baselines, despite we design a fine-grained reconstruction process. Across all applications and update levels, its reconstruction latency has an average relative difference of XX to XX seconds versus baselines. Leveraging transmission optimization, \toolName achieves XX\% up to end-to-end latency reduction over the baselines.
% \toolName introduces onboard update overhead comparable to the baselines, despite employing a fine-grained reconstruction process. It increases only an average difference of by about XX seconds compared to baselines.
Despite employing a fine-grained reconstruction process, \toolName incurs onboard update overhead comparable to the baselines, introducing only an average increase of about 2 seconds. It achieves 100\% update correctness across all evaluated applications.
% By leveraging transmission optimizations, \toolName achieves up to XX\% reduction in end-to-end latency compared to the baselines.
}
\vspace{-6mm}

\subsection{RQ3 Results (Recovery Overhead)}\label{sec:rq3faulttolerant}

% \begin{table}[t]
% \footnotesize
%  \caption{A comparative analysis of recovery mechanisms.}
%  \vspace{-3mm}
%     \label{tab:RQ3recovery}
%     \begin{tabular}{c|c|c|c}
%     \hline
%     \begin{tabular}[c]{@{}c@{}}\textbf{Recovery}\\ \textbf{Mechanisms}\end{tabular} & \begin{tabular}[c]{@{}c@{}}\textbf{Storage}\\ \textbf{Size (MB)}\end{tabular} &  \begin{tabular}[c]{@{}c@{}}\textbf{Backup}\\ \textbf{Time (s)}\end{tabular} & \begin{tabular}[c]{@{}c@{}}\textbf{Recovery}\\ \textbf{Processing Time (s)}\end{tabular}  \\
%     \hline
%     Image Recovery &965325824B& X & X \\
%     \hline
%     File Recovery &84726B&3331.36ms&43879.51322ms  \\
%     \hline
%     Patch Recovery &6655B&3320.74ms&43845.99092ms   \\
%     \hline
%     \textit{\textbf{Ours}} &193377780B&0.479645199ms&36131.41073ms \\
%     \hline
    
% \end{tabular}
% \vspace{-4mm}
% \end{table}

\begin{table}[t]
\footnotesize
 \caption{A comparative analysis of recovery mechanisms.}
 % \vspace{-3mm}
    \label{tab:RQ3recovery}
    \begin{tabular}{c|r|r|r}
    \hline
    \begin{tabular}[c]{@{}c@{}}\textbf{Recovery}\\ \textbf{Mechanisms}\end{tabular} & \multicolumn{1}{c|}{\textbf{Storage Size}}  & \multicolumn{1}{c|}{\textbf{Backup Time}}  & \begin{tabular}[c]{@{}c@{}}\textbf{Recovery}\\ \textbf{Processing Time}\end{tabular}  \\
    \hline
    Image Recovery &942,701.00 KB &  \multicolumn{1}{c|}{--} & \multicolumn{1}{c}{--} \\
    \hline
    File Recovery &82.74 KB & 3,331.36 ms & 43,879.51 ms  \\
    \hline
    Patch Recovery &6.50 KB & 3,320.74 ms & 43,845.99 ms   \\
    \hline
    \textit{\textbf{Ours}} & 188,845.49 KB & \textbf{0.48 ms} &36,131.41 ms \\
    \hline
    
\end{tabular}
\vspace{-3mm}
\end{table}

The recovery mechanism of \toolName is a layer-based rollback, which preserves only the previous application layer and performs quick revision. Leveraging the original application version and fine-grained upload package available onboard, we compare it with three alternative recovery strategies: (1) Image recovery, which retains a complete container image of the previous version and restores it directly without additional computation; (2) File recovery, which reconstructs the prior version by identifying and reversing file-level changes introduced during the update; and (3) Patch recovery, which utilizes recorded reverse edit paths to perform fine-grained recovery. To evaluate their recovery capability in the context of update failures, we analyze their recovery overhead using the satellite application. Recovery overhead consists of two parts: backup time, the duration required to prepare necessary recovery data, and recovery processing time, the time needed to restore the application after a failure. For this evaluation, App1 with a 10\% code modification level is selected as the evaluation case.

As shown in Table~\ref{tab:RQ3recovery}, our layer rollback of \toolName strikes an effective balance. Specifically, image recovery does not have recovery overhead, but at the cost of high storage demand (942,701.00 KB), which is infeasible in resource-constrained satellites. File and patch recoveries significantly reduce storage usage (82.74 KB and 6.50 K), but suffer from long backup time (over 3,300 ms) and high recovery processing time (over 43,800 ms). Thus, file and patch recoveries are prone to recovery failure, due to the extended duration required for their backup phases.
% Moreover, its recovery succeeds only if the failure occurs after the backup completes. Due to the long backup time, patch-based recovery yields a low hit rate (XX\%).
% Moreover, its recovery success depends on whether failures occur after the backup is complete. Based on the long backup time, patch-based recovery has a low hit rate (XX\%).
% For their stability, file and patch recoveries achieve low hit rates of XX\% and XX\%, respectively, due to several failures occurring during its long backup phase. 
In contrast, our layer recovery uses moderate storage while achieving ultra-fast backup (0.48 ms) and reduced recovery processing time (36,131.41 ms), outperforming file- and patch-based methods in responsiveness and reliability. The efficiency stems from the immediate preservation of only the relevant application layer, enabling timely rollback for most failures.
% and achieves fast backup and recovery (backup: 0.48 ms, recovery: 36,131.41 ms). It also demonstrates a lower risk of recovery failure than file and patch recoveries, it allows for quick retention (only 0.48 ms, significantly faster than the other two methods) of partial preservation of the old layer during the update process, enabling timely rollback for most failure events. 
Overall, \toolName provides a practical and efficient recovery solution that avoids the high storage of image recovery and the computational burden and unreliability of file and patch recoveries. 

% Thus, our recovery solution is suitable for satellite scenarios with resource constraints.

\vspace{-2mm}
\finding{
\toolName achieves a balanced and efficient fault-tolerant recovery mechanism.
}
\vspace{-6mm}

\subsection{RQ4 Results (Real-World Case)}

% To validate the practicability of \toolName on onboard application update, we conduct a case study on a real-world on-orbit satellite, called \wjf{XXX}. This satellite is equipped with the ability to communicate with ground stations and has the necessary cloud-native computing power on board. We use the data encryption task commonly used on satellites~\cite{} to assess the practicability of onboard application reconstruction and update of \toolName.

% To validate the practical applicability of \toolName for real-world on-orbit deployment, 
% In this section, we conduct a real-world case study on an operational LEO satellite with onboard cloud-native computing capabilities. The satellite supports container-based application deployment and remote communication with ground stations, providing a real-world environment to evaluate the update process of \toolName.

This section presents a real-world case study on an operational LEO satellite called BUPT-2 equipped with onboard cloud-native computing. BUPT-2 supports containerized application deployment and remote communication with ground stations, offering a real-world environment to initially verify the practicability of the onboard update process of \toolName.

Due to the limited time we have available to experiment with this satellite, we conduct only one test case. We select a data encryption task, commonly used on satellites~\cite{ingemarsson2003encryption, banu2009fault}. Its initial version is containerized, uploaded, and deployed on the satellite. 
% its difference update content is prepared on the ground in the form of an upload package, capturing only the code changes. Along with this package, the application reconstruction script implementing \toolName's onboard application update as well as the related satellite control script are uploaded to the satellite. Once delivered, the satellite executes the reconstruction script to perform the fine-grained application reconstruction process entirely onboard. After reconstruction, the satellite performs a content check to the functionality correctness of application update. The test results validate the practicality of \toolName for onboard satellite application updates in satellite scenarios.
Subsequently, a corresponding upload package, containing the differential code, is prepared on the ground. Along with this package, we upload the application reconstruction script implementing \toolName's onboard update strategy, as well as the necessary control script. Upon receipt, the satellite executes the reconstruction script to perform a fine-grained application reconstruction process for the data encryption task entirely onboard. During in-orbit testing, we collaborate with satellite operators to collect and analyze telemetry data. The ``successCount'' field, tracking successfully executed remote control commands, served as an indirect yet reliable indicator of application correctness and executability post-update. Furthermore, operators explicitly confirm the update's success and verify normal satellite system operation. The results demonstrate the practicality of \toolName for satellite application updates in real satellite environments.

% \wjf{In addition, since we applied \toolName to BUPT-2 satellite, we incorporated feedback from the practitioners who developed and operated the satellite. Their insights help validate \toolName’s real-world applicability beyond technical metrics. During the in-orbit testing of \toolName, we collaborated with satellite operators and collected relevant telemetry data. Specifically, we analyzed the "successCount" field, which records the number of successfully executed remote control commands, as an indirect but reliable indicator of the updated application’s correctness and executability. Furthermore, satellite operators explicitly confirmed the success of the software update and verified that the satellite system operated as expected post-update.}

\vspace{-2mm}
\finding{
% \toolName has practical feasibility for satellite application updates in real-world on-orbit satellite deployment.
\toolName shows the practicality of updating satellite applications on real-world in-orbit satellites.
}
\vspace{-6mm}

\section{Threats to validity}\label{sec:discussion}

\noindent \textbf{Measurement Bias.} Performance evaluation, including transmission latency and onboard update latency, may be influenced by communication variability and onboard resource fluctuations. To reduce this, we assume a constant uplink bandwidth, as satellite S-band communication is stable and predictable~\cite{SBand, zhang2024resource}, making transmission latency effectively deterministic. For onboard update latency, we report the mean over multiple runs~\cite{liu2023faaslight, wang2018peeking} as the final result of each experiment to reduce the impact of transient variations and improve result reliability.

% In our paper, we measure the transmission latency and onboard update latency for satellite applications. Since these evaluation metrics may be affected by communication or the current resource ability, the obtained latencies may lead to possible bias. To alleviate this threat, for the calculation of transmission latency, we use a constant uplink bandwidth. This is because the uplink leverages the low-frequency S-band to communicate~\cite{SBand}. Thus, transmission latency is deterministic and is not affected by potential factors. For onboard update latency, we follow the common practice, which calculates the average latency of multiple runs to represent the general level of the current experiment. Thus, we report the average result for multiple repetitions.

% \noindent \textbf{Conclusion Generalizability.} The conclusions of \toolName and baseline methods may vary across different types and language implementations of satellite applications, potentially impacting conclusion generalizability. To mitigate this, we evaluate 10 representative applications spanning diverse satellite tasks, including image object detection and tracking, core network services, and image encoding. Moreover, these applications are implemented in diverse languages, such as Python, C, Go, C++, MATLAB, JavaScript, and mixed languages, reflecting real-world heterogeneity. Therefore, the conclusions drawn are based on broad empirical evidence and are reasonably generalizable across satellite applications. 
\noindent \textbf{Conclusion Generalizability.} The generalizability of conclusions drawn from \toolName and baselines may be influenced by variations in satellite application types and languages. To address this, we evaluate representative applications covering a broad spectrum of satellite tasks, e.g., object detection and tracking, core network services, and image encoding. These applications are implemented in diverse languages, e.g., Python, C, Go, C++, JavaScript, and mixed languages, thereby reflecting the application heterogeneity. Thus, our conclusions are grounded in diverse empirical evidence and are reasonably generalizable across satellite application scenarios.
% Moreover, we validate the practical applicability of \toolName through a real-world validation on an operational in-orbit satellite.

% \textbf{Conclusion effectiveness.} We evaluate the effectiveness of \toolName and other baselines on satellite applications. Technical effectiveness may vary depending on the task type and languages of satellite applications. This may potentially influence the experimental conclusions that we summarize herein. To mitigate it, we investigate 10 satellite applications covering a variety of satellite tasks, e.g., image object detection, core network functionality, image change detection, and encoding, as well as object tracking. In addition, the satellite applications we collect from top-tier satellite-related papers are written in different programming languages, e.g., Python, C, Go, C++, Matlab, and JavaScript, as well as mixed languages. Thus, the conclusion of technical effectiveness is based on testing results for various types and languages of satellite applications, thereby the technical effectiveness conclusion is generalizable and reliable.

\section{Related Work}\label{sec:relatedwork}

\noindent \textbf{Satellite Computing}
Recent advances in satellite computing have focused on enhancing onboard computational capabilities to support increasingly complex satellite applications. Komet~\cite{pfandzelter2024komet} was a serverless edge computing platform tailored for LEO satellites, enabling on-demand function execution and dynamic service migration through hardware abstraction. Xing~\textit{et al.}~\cite{xing2024deciphering} analyzed how thermal control and power constraints affect task scheduling on COTS devices in orbit. SECO~\cite{zhai2024seco} was a collaborative edge computing framework to coordinate real-time tasks, such as image compression and object detection, across multiple satellites for low-latency Earth observation. Phoenix~\cite{liu2024orbit} was an energy-ware scheduling framework that leveraged sunlit orbital periods to opportunistically perform onboard processing. To alleviate bandwidth constraints, Maskey and Cho~\cite{maskey2020cubesatnet} designed a lightweight convolutional neural network for onboard image classification, reducing unnecessary data transmission. The Tiansuan constellation~\cite{wang2021tiansuan} employed KubeEdge to enable satellite-ground collaborative inference, improving application responsiveness. While these efforts advance various aspects of satellite computing, little attention has been given to the approaches for updating and evolving satellite applications. In this work, we address this gap by presenting \toolName.

\noindent \textbf{Software Update}
Research on software updates has been categorized into three primary phases: pre-update, in-update, and post-update. Prior research on pre- and post-update phases has primarily focused on update planning and impact assessment. In the pre-update stage, efforts have centered on recommending \textit{what} to update. Techniques include mining user feedback~\cite{di2016would, zhou2020user}, recommending configuration updates~\cite{hassan2018rudsea}, and leveraging large language models for code change suggestions~\cite{liu2025automatically}. However, these approaches focus on update recommendation rather than execution. Most assume a ground-based context, where direct version replacement is feasible, ignoring the operational constraints present in resource-limited satellites.
In the post-update stage, studies have addressed user behavior and update effectiveness. This includes analyzing the influence of automated pull requests~\cite{mirhosseini2017can}, detecting faulty updates~\cite{saidani2022tracking}, evaluating update interval strategies~\cite{di2022software}, and characterizing update patterns~\cite{kotzias2019mind}. While valuable, these studies address different concerns and do not tackle the unique constraints of software evolution in satellites.

Studies on in-update have explored dynamically how to apply updates safely and efficiently, i.e., dynamic software update (DSU). For instance, Zhao~\textit{et al.}\cite{zhao2014automated} leveraged test cases to identify safe update points in Java programs, albeit with high computational cost. Cazzola and Jalili\cite{cazzola2016dodging} proposed an annotation-driven framework to detect unsafe update points and dynamically avoid them. MCR~\cite{giuffrida2016automating} supported live updates in C applications by automating control transfer and state migration through static instrumentation and thread coordination. 
However, DSU techniques pursue incremental, in-place updates during program execution, fundamentally differing from our goal of stable application replacement on satellites, which are infrequent, planned, communication-limited, and ground-controlled. Additionally, DSU relies on language-specific analysis, runtime monitoring, and extensive test suites, making it impractical for resource-constrained satellites. In contrast, \toolName employs a lightweight content-aware update framework that avoids such overhead and addresses unique challenges to support diverse satellite applications.

\section{Conclusion}\label{sec:conclusion}

% This paper presented \toolName, an effective satellite application update framework tailored for satellite computing. \toolName tackled the unique constraints of satellite environments by supporting content-aware difference updates, efficient onboard application reconstruction, and fault-tolerant recovery. Satellite simulation environment evaluation across diverse satellite applications demonstrates the effectiveness of \toolName, achieving significant reductions in transmission latency (on average XX\%) and end-to-end update time (on average XX\%). Additionally, we validate its practicability through real-world deployment on an operational in-orbit satellite, confirming its suitability for real-world satellite software maintenance and evolution.

This paper presented \toolName, an effective framework for updating satellite applications under the constraints of satellite computing. \toolName adopted a container-based design for heterogeneous applications. Moreover, it enabled content-aware differential updates, onboard application reconstruction, and fault-tolerant recovery. Evaluation in a satellite simulation shows that \toolName significantly reduces transmission latency by up to 91.18\%, with an average reduction of 56.54\%, compared to the best-performing baseline. It also ensures 100\% correctness in application updates. A real-world deployment on an operational in-orbit satellite confirms the applicability of \toolName for satellite software maintenance and evolution.

\section*{Acknowledgment}
This work is supported by the China Postdoctoral Science Foundation under Grant No. 2025M771560, Postdoctoral Fellowship Program (Grade B) of China Postdoctoral Science Foundation under Grant No. GZB20250400, National Natural Science Foundation of China under Grant Nos. U21B2016, 62425203, and 62032003, and Fundamental Research Funds for the Central Universities under Grant No. 2024ZCJH11.

% Beijing Natural Science Foundation under Grant No. 4244076, National Natural Science Foundation of China under Grant Nos. 62425203 and 62032003, and Beijing Postdoctoral Research Foundation under Grant No. 2024-ZZ-20.

\bibliographystyle{IEEEtran}
\bibliography{satellite}

\end{document}